# Merowe Dam and the inundation of paleochannels of the Nile

Amelia Carolina Sparavigna
Dipartimento di Fisica, Politecnico di Torino
Corso Duca degli Abruzzi 24, Torino, Italy

**Abstract**: The course of the Nile in northern Sudan follows a contorted path through bedrocks, creating the Great Bend. Few years ago, the satellite images showed a fertile strip of land with villages, where paleochannels of the river hosted many fields with cultivations and archaeological sites. Now, a huge part of this valley is under the waters of Merowe Dam reservoir. Comparing the images of the region before and after the dam gates were closed, we can see that the reservoir created itself through flooding the paleochannels.

The huge reservoir of the Merowe Dam on Nile River is featured in an image photographed on 5 October 2010 by a crewmember on the International Space Station [1]. The dam is located near the Fourth Cataract of the river, in that part of Nubia desert where the Nile is creating its Great Bend, a contorted path of the river through the bedrocks of Bayuda Massif. Few years ago, the satellite imagery showed a fertile strip of land with many villages, where the paleochannels of the Nile hosted vivid green cultivated fields. With term "paleochannel", we call the old dry bed of rivers. Now, a huge part of this valley is under the waters of Merowe Dam reservoir that flooded villages, fields and archaeological sites. The dam was built to generate hydroelectric power, intended to help industrial and agricultural development of the country.

The upper panel of Figure 1 is reproducing the NASA image, which illustrates the current extent of the water reservoir behind the dam. All the gates were closed in 2008. In the lower part of the image, we can see for a comparison, on the left, a part of the region before the flood as found in Google Maps imagery and, on the right, the same area as shown by NASA images, after rotation, size adjustment and enhancement with a suitable image processing. The aim of this paper is a discussion on the role of paleochannels in the development of this dam reservoir, as can be obtained from the satellite imagery. However, before start discussing, let us remember the framework of the building of this dam and the total project where the Merowe Dam is only a part.

Following Sudan's independence from Egypt and the United Kingdom in 1956, allocation and control of Nile River water was divided between Egypt and Sudan by the Agreement for the Full Utilization of the Nile Waters (Nile Waters Treaty), signed on 8 November 1959 [2]. In May of 1999, the Nile Basin Initiative (NBI) was launched, joining all nations of the basin. The objectives for the NBI (see [3]) include the development of water resources of the Nile in an equitable way, to ensure efficient water management, cooperation and joint actions. In May 2004, the "Nile Transboundary Environmental Action Project", the first of eight basin-wide projects under the NBI, was launched in Sudan. Nile River diversion and work on the dam began in early 2004. The work finished when the water level in the reservoir reached 300 m above mean sea level in 2009. The dam was inaugurated on March 3, 2009 [4].

Merowe Dam is the second huge construction, after the Aswan Dam, obstructing the waters of the Nile [5-8]. Before the Aswan Dam was built, the river flooded the Egypt each year during late summer, with the waters flowing down its African drainage basin. The floods brought natural nutrients and minerals, annually enriching the soil and rendering the lands along the Nile an ideal area for farming. Egypt developed the dam to control the floods, to protect and support farmland and the cotton production, quite relevant for the country. However, damming the Nile caused a number of environmental and cultural problems. It flooded much of Lower Nubia [7], and 50,000 people were displaced [8]. The silt, which the Nile deposited in the yearly floods, making the Nile floodplain fertile, is now held behind the dam. Silt deposited in the reservoir is lowering the water storage capacity of Lake Nasser. Ref.5 lists several facts about erosion and others problems connected with the dam and the huge damages that concerned the archaeological sites. The rescue of some of them began in 1960 under UNESCO. The Great Temple of Abu Simbel was relocated on the shores of Lake Nasser [7]. Other monuments were granted to countries that helped with the works. Nevertheless, many archaeological sites have been flooded by Lake Nasser.

The creation of Merowe Dam constrained again several local settlements to be displaced in desert regions [9,10]. Moreover the water covered a region which contains significant but little-studied archaeological sites [4,11]. Downstream the dam, there is Napata, one of the archaeological sites of Nubia (on Nubia, see references [12-19]); this site is not directly threatened, but the local climate change could create problems to the preservation of the monuments (see Fig.2 for a map). A variety of international institutions conducted archaeological surveys to gain as much information as possible from sites that will be destroyed or otherwise made inaccessible due to the flood [20,21]. As told in Wikipedia, very little archaeological work has ever been conducted in this particular region [4]. The item is telling that "surveys have confirmed the richness and diversity of traceable remains, from the Stone Age to the Islamic period" and that the "main problems are the shortness of the remaining time and limited funding. Unlike the large UNESCO campaign conducted in Egypt before the completion of the Aswan High Dam, when more than a thousand archaeological sites … were moved to prevent them from drowning …, work at the Fourth Cataract is much more restricted" [4].

The problems of displacement of people and lost of archaeological sites, already encountered for Aswan and Merowe dams, will be present again in next times, because the Sudanese government is planning to build other three dams. The dams are: Kajabar at the Third Cataract Nile, Saiteet on Atbara River and Al-Sheraik at the Fifth Cataract Nile, northern of Atbara city [22]. This project is then involving the archaeological Nubian regions of Kerma and Meroe. Let us remember that archaeological evidence has confirmed that Nubia, was inhabited at least 70,000 years ago and that a settled culture appeared around 8,000 B.C [13-14]. By the 5th millennium BC, the people who inhabited what is now called Nubia participated to the so-called Neolithic revolution [15,16]. In the following, Nubia hosted the Kingdoms of Kerma and Kush, having strong connections with Egypt [17-18,23].

Figure 2 is showing the locations of two famous archaeological sites (Napata and Meroe, purple markers), besides that of Merowe Dam itself. The map was obtained with ACME Mapper. The red markers near the Nile indicate the distribution of villages, for that part of the river valley where Google imagery has high resolution enough for a localization of possible archaeological remains. The blurring of image in Fig.2 represents the area where maps have low resolution. Near these settlements, there are places where the ground is covered by small mounds, round with flattop or with a depression as a hole, probably burial sites [11]. The archaeological sites are generally located in paleochannels, that is, the old dry beds of the branches of Nile river, as can be deduced from the satellite maps after an image processing able to enhance these dry channels. In this part of the Nile Valley, the strip of land suitable for settlements is so narrow that the paleochannels supported and are supporting cultivated fields.

The fact that remote sensing, for instance by means of satellites equipped with radar sensors, is able to see the channels buried under the sand is well known. In fact, a huge paleochannel of Nile was discovered with a SIR-C/X-SAR imaging radar system [24] during two flights of the NASA space shuttle Endeavor in 1994. These data revealed how the bedrock structures of different age control much of the Nile's course. More recently, a Shuttle Radar Topographic Mission (SRTM) revealed segments of an inactive drainage channel systems in eastern Sahara [25]. One of the examined regions is the border region between Egypt and Sudan. As reported in [25], SRTM analysed the drainage system under desert-sandy surfaces.

It could be surprising, but Google Imagery shows the dry channels of the ancient drainage systems as well as if we were observing them with SAR data. We discussed in Ref.[26], that the paleochannel of Nile, the same analysed in [24], is clearly displayed with Google Maps, after an image processing with a method based on free software: AstroFracTool [27] and Iris.

Let me then consider the reservoir of Merowe Dam and the aim of this paper again, that is, the comparison of ancient drainage systems of Nile with the current reservoir extension. As previously told, we use the image obtained from NASA, reproduced in the upper part of Figure 1, and images from Google Maps, which is showing the region when the dam was under construction. To have a proper comparison between images, the original image from NASA is rotated and resized with GIMP, the GNU Image Manipulation Program. To enhance the details, as in Ref.26 and 27, AstroFracTool and Iris software have been used on NASA and Google Maps imagery.

In Figure 3, we can see the two images from the lower part of Figure 1, used as two levels merged with GIMP, to obtain a superposition and have then a better comparison. Note that many dendritic drainage systems appear clearly inundated by the reservoir of Merowe Dam. Figure 4 shows in more detail one of these dendritic structures. The upper/left image is a Google Maps image after processing. The lower/right image is the NASA one after being rotated, resized and image-processed. The other two images are giving a result with different percentage of superposition. Note that Google clearly displays a paleochannel: the water inundated this paleochannel as NASA image is evidencing.

Has the dam reservoir produced effects on the huge paleochannel of the Nile discovered by the Shuttle X-SAR mission [24,26]? Again, we can compare a Google image showing this structure with the corresponding detail obtained from the NASA image of Fig.1. The answer to our question is positive: we see a branch of the reservoir moving into the old channel. A further increase of the water level could provoke the inundation of this old and dry bed of the Nile.

To the author's knowledge, this paper is the first one proposing a comparison of images before and after the creation of the reservoir, to evidence the inundation of paleochannels. Studies on the inundation processes produced by dam reservoirs are fundamental, first for the safety of people and animals, and for conservation of monuments and archaeological and historical places. It is author's opinion that the use of satellite images, in particular of the free service connected with Google Maps, can help improving models and simulations when paleochannels are involved. Let me conclude that in the future we will probably see the water flowing in the paleochannel of Figure 5 as in a revitalized river. May be the readers could consider this fact quite hard to accept. For this reason, the author invites the reader to look at Egypt with Google Maps at the following coordinates: latitude 22.7958, longitude 31.47514.

commonly refers to the High Dam, which is the larger and recent one, planned in 1954 and constructed between 1960 and 1970, after Egypt's independence from the United Kingdom. The Old Aswan Dam, or Low Dam, was created during the British colonial period. After 1882 occupation of Egypt, British Empire built it from1898 until 1902. Despite initial limitations imposed on its height, due to concern for the Philae Temple, the initial height was raised twice. The dam has now a level at 36 m above the original riverbed. With the construction of the High Dam upstream, the Old Dam's ability to pass the flood's sediments was lost. The Old Dam reservoir level was also lowered and now the dam provides controlling the waters coming for the High Dam. Immediately after the planning for the High Dam, archaeologists began raising concerns that several historical and archaeological sites were about to be covered by water. The rescue of some of them began in 1960 under UNESCO. In 1958, Soviet Union provided funding for the dam project, in 1960, it started the construction and in 1970 the High Dam was completed.

05-06.html

[23] The first kingdom that unified much of Nubian region was the Kingdom of Kerma. Named for its presumed capital at Kerma, this kingdom was one of the earliest urban centers in Sub-Saharan Africa. By 1750 BC, the kings of Kerma were powerful enough to create monumental structures and have rich tombs with possessions for the afterlife and large horrifying human sacrifices. At one point, Kerma came very close to conquering Egypt. When Egyptian power revived under the New Kingdom (c.1532–1070 BC), Egypt began to expand southwards. Destroying the kingdom of Kerma, Egypt expanded to the Fourth Cataract. By the end of the reign of Thutmose I in 1520 BC, all of northern Nubia had been annexed. They built an administrative center at Napata. When the Egyptians pulled out, they left their legacy merged with indigenous customs that formed the kingdom of Kush. In the 8th century BC the Kush kingdom, capital Napata, gradually extended its influence into Egypt. About 750 BC, a Kushite king conquered the Upper Egypt and became the ruler of Thebes. His successor conquered the Delta, uniting Kush and Egypt under the XXV Dynasty, for about a hundred years. Under the pressure of Assyrians, the last Kushite pharaoh, Taharqa (688–663 BC), withdrew and returned to Napata, where his dynasty continued to rule Kush and extended dominions to south and east. In 590 BC, an Egyptian army sacked Napata, compelling the Kushite court to move to Meroe near the Sixth Cataract. The Meroitic kingdom developed independently of Egypt, and during the height of its power in the 2nd and 3rd centuries BC, Meroe extended over a region from the Third Cataract till the gates of Khartoum. About 350 AD, an Axumite army from Abyssinia captured and destroyed Meroe city, ending the kingdom's independent existence.

[24] The Origin of the Great Bend of the Nile from SIR-C/X-SAR Imagery, R.J. Stern, M.G. Abdelsalam, Science, 1996, Vol. 274(5293), pp. 1696-1698,

[25] Radar topography data reveal drainage relics in the eastern Sahara, E. Ghoneim, C. Robinson, F. El-Baz, Int. J. of Remote Sensing, 2007, Vol.28(8), doi:10.1080/01431160600639727

[26] Enhancing the Google imagery using a wavelet filter, A.C. Sparavigna, Geophysics (physics.geo-ph); Earth and Planetary Astrophysics (astro-ph.EP), 2010, arXiv:1009.1590

[27] Crater-like landform in Bayuda desert (a processing of satellite images), A.C. Sparavigna, Geophysics (physics.geo-ph); Earth and Planetary Astrophysics (astro-ph.EP), 2010, arXiv:1008.0500

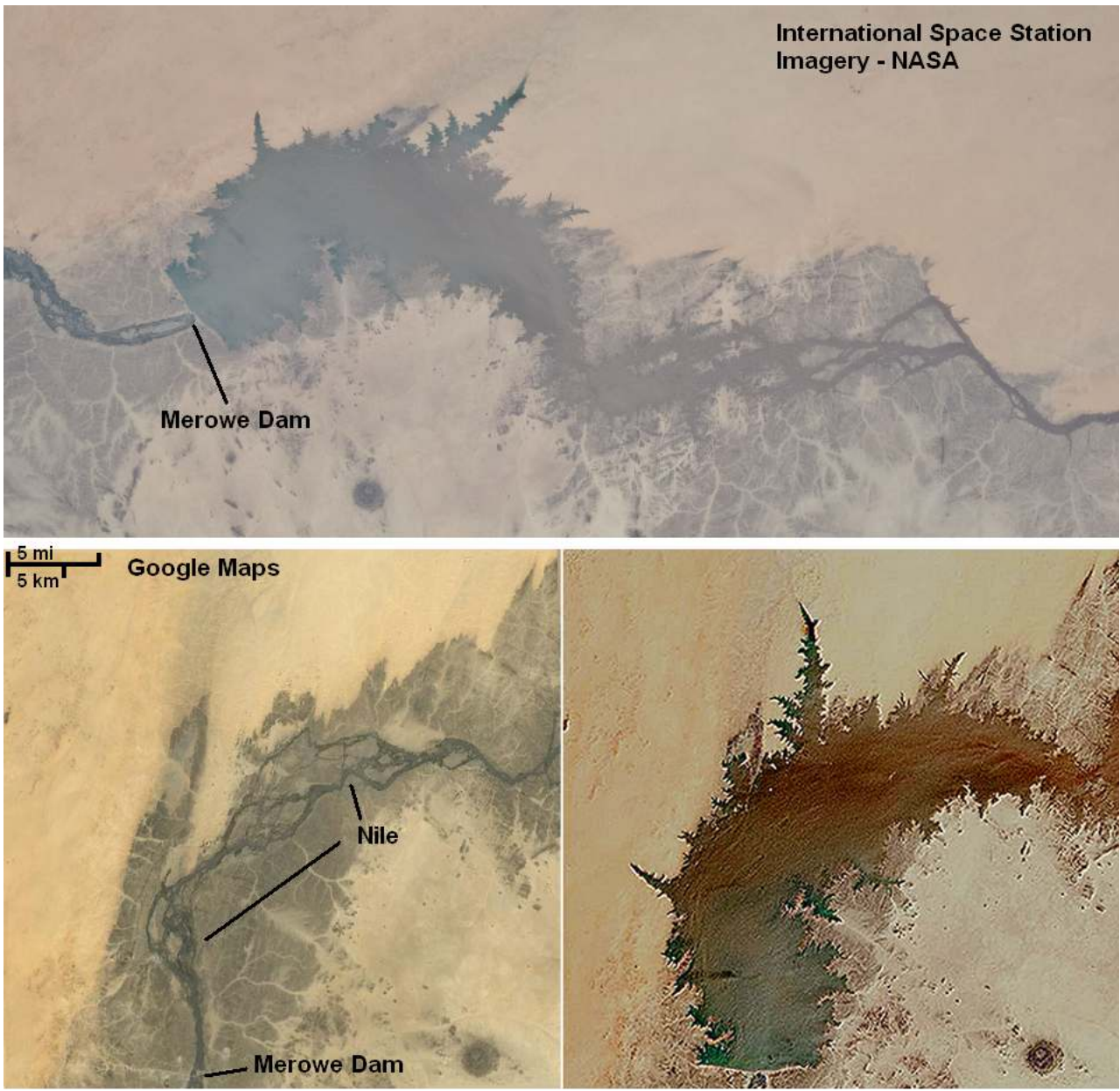

Figure 1: The reservoir of Merowe Dam on river Nile as photographed on 5 October 2010 by a crewmember on the International Space Station, NASA. The dam is located near the Fourth Cataract of the river. In the lower part of the figure, on the left, the region before the inundation as found from Google Maps and, on the right, the same area as shown by the NASA image, after enhancing with a suitable image processing for comparison.

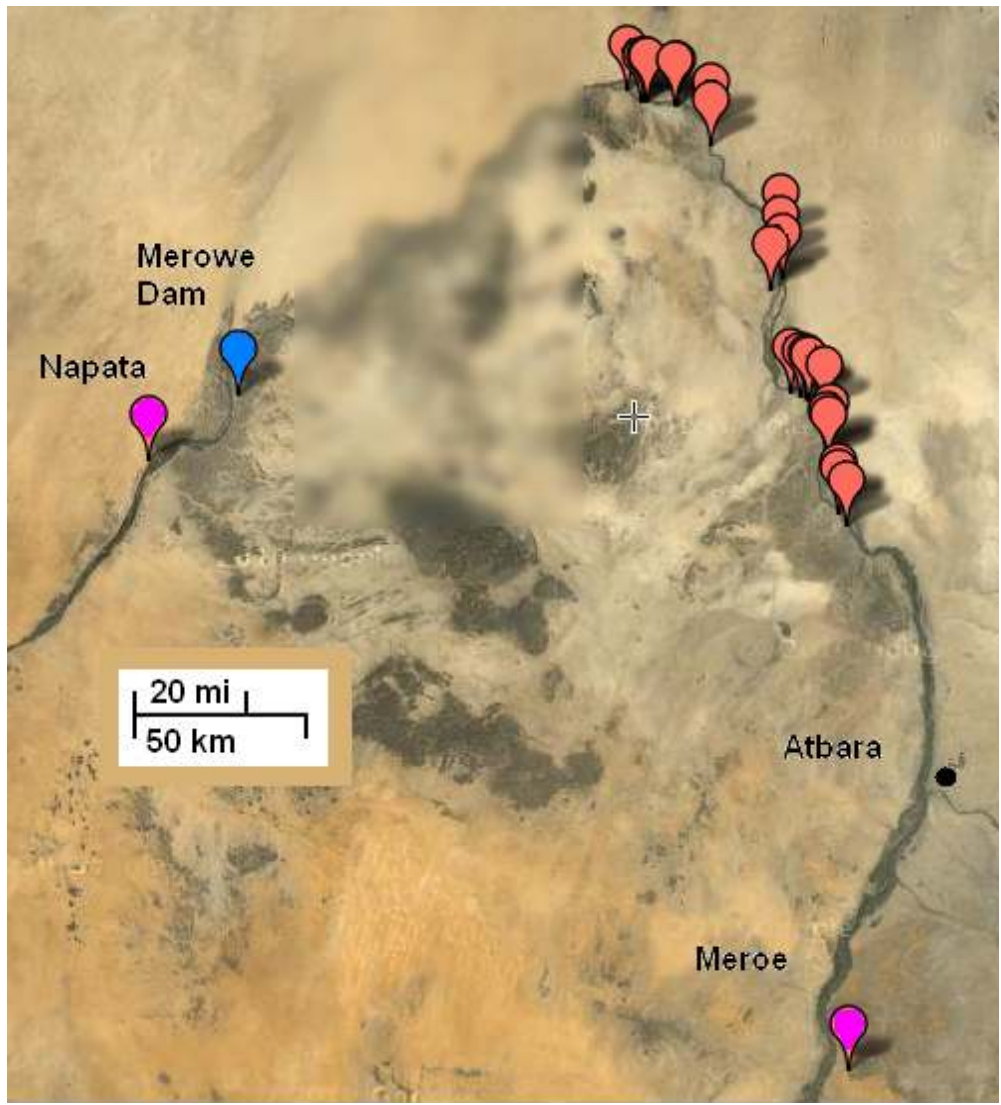

Figure 2: This image is showing the locations of the dam (blue marker) and of two famous archaeological sites (Napata and Meroe, purple markers). The map was obtained with ACME Mapper. The red markers indicate the distribution of villages in that part of the river valley where Google imagery has a resolution suitable to localize possible archaeological remains (the blurring represents the areas where maps have low resolution). Near these settlements, there are areas where the ground is covered by small mounds, round with flattop or with a depression as a hole, probably burial sites (see http://arxiv.org/ftp/arxiv/papers/1011/1011.3716.pdf). The archaeological sites are generally located in paleochannels of the Nile, that is, old dry beds of branches of the river.

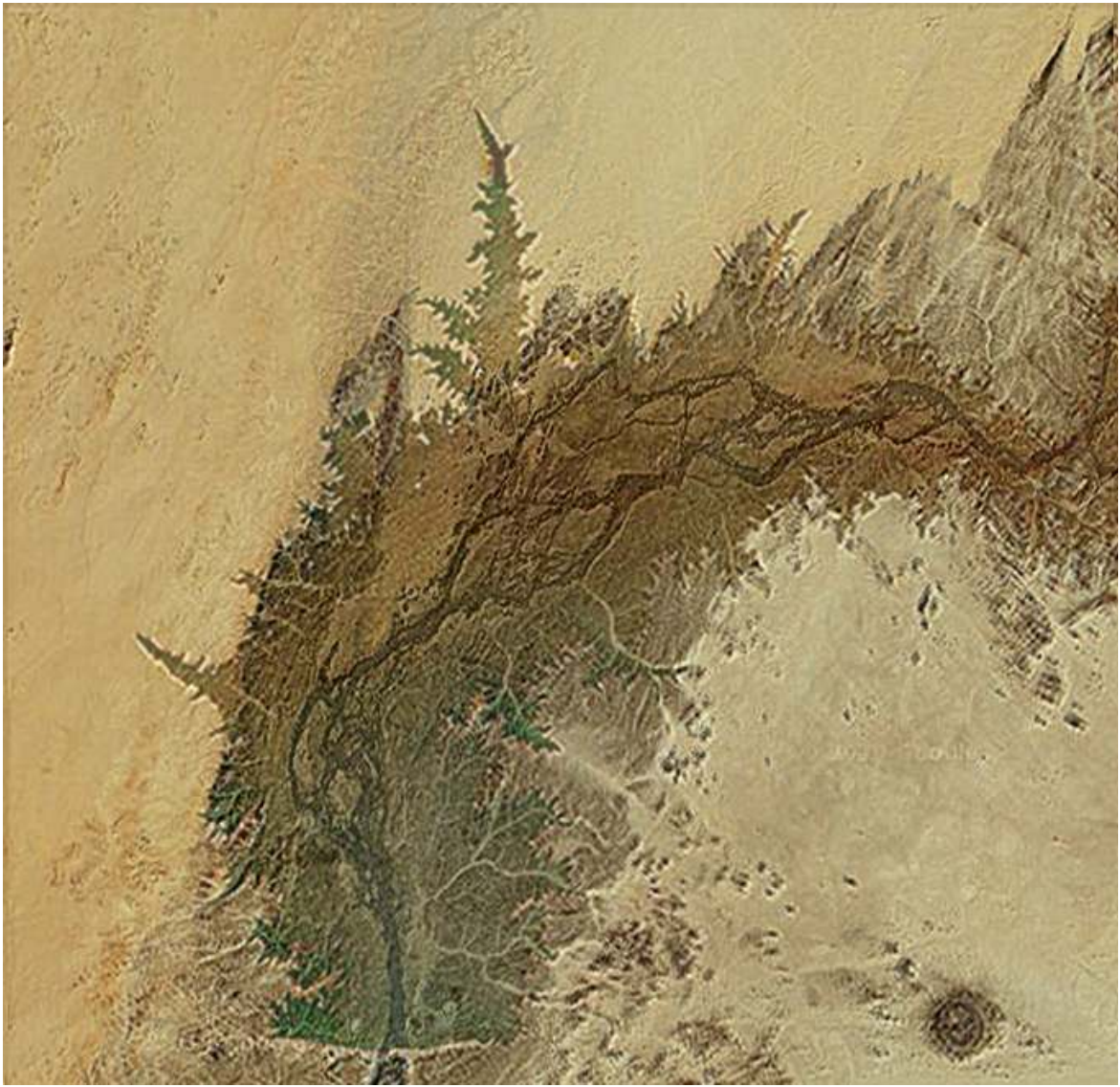

Figure 3: The two images from the lower part of Fig.1 after superposition, to have a better comparison. Note the dendritic drainage systems inundated by the reservoir of Merowe Dam.

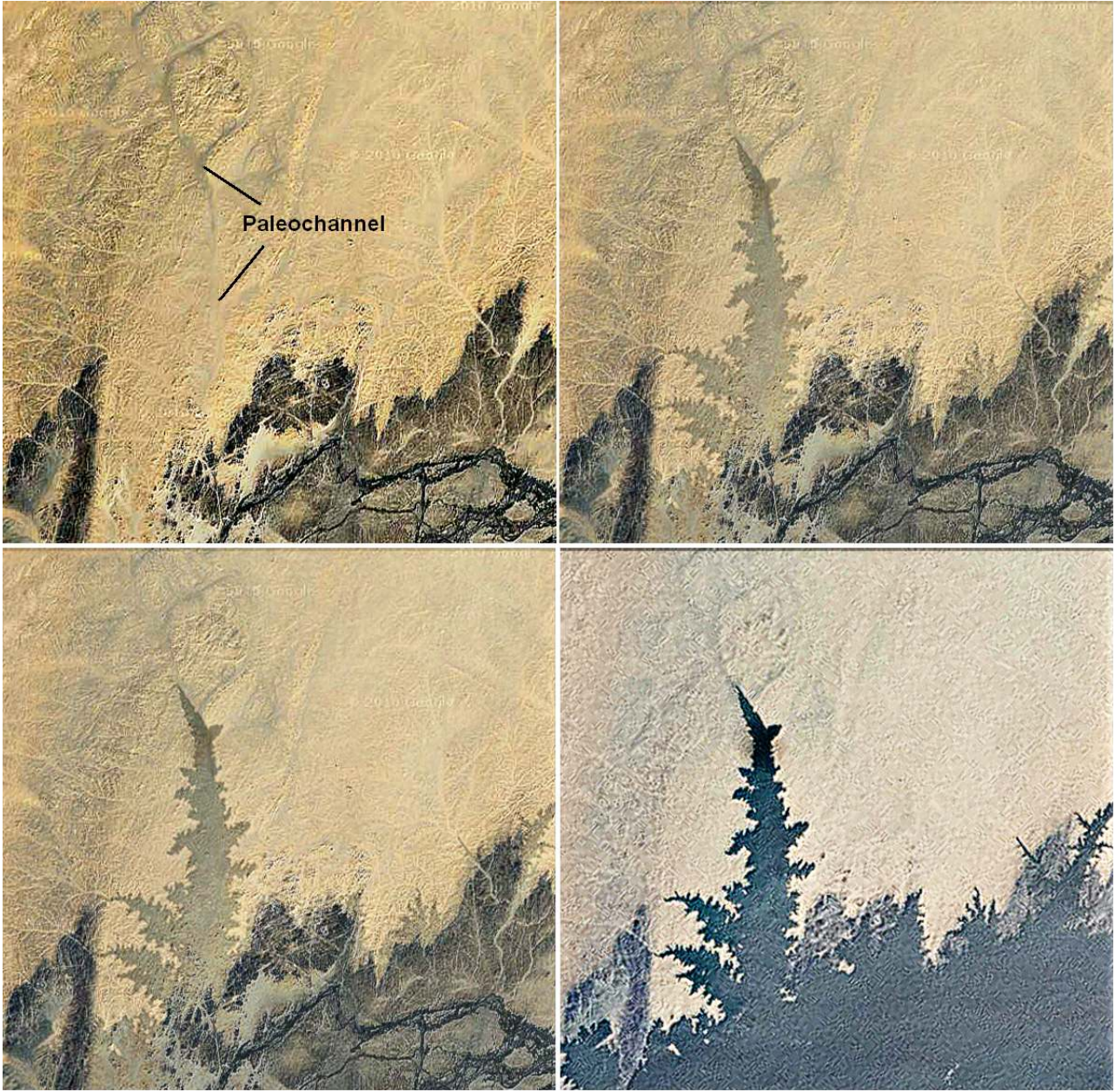

Figure 4: The image shows in detail one of the dendritic structures of the Merowe Dam reservoir. The upper/left image is coming from Google Maps after processing as in Ref.26. The lower/right image is the NASA one after being rotated, resized and image-processed. The other two images are giving the result with different percentages of superposition. Note that in the upper/left one the paleochannel is clearly displayed. The water inundated it as NASA evidenced.

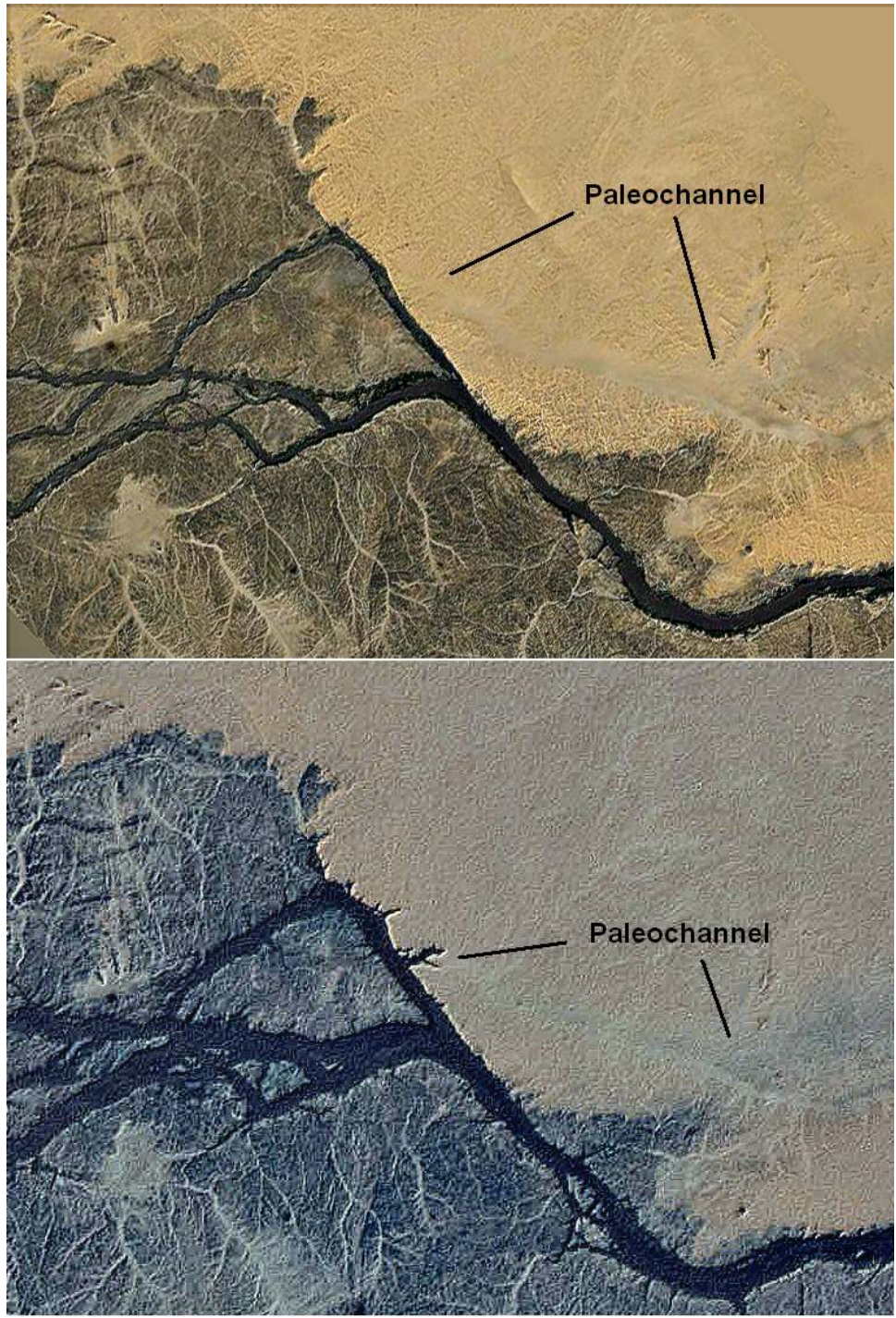

Figure 5: The reservoir is starting to affect the huge paleochannel of the Nile discovered by the Shuttle X-SAR mission [24,26]. The upper panel shows an image obtained from Google after a proper enhancement. The lower panel shows the same structure as obtained after rotation, resize and enhancement of a detail from the NASA image of Fig.1.